\documentclass[fleqn,10pt]{wlscirep}
\usepackage[utf8]{inputenc}
\usepackage[T1]{fontenc}
\usepackage{xspace}
\newcommand\wmsphi{\ensuremath{\Phi^{\mathrm{WMS}}}\xspace}
\newcommand\phiR{\ensuremath{\Phi^{\mathrm{R}}}\xspace}
\title{Quantifying Hierarchical Selection}

\author[1,2,3*]{Hardik Rajpal}
\author[4]{Clem von Stengel}
\author[1,5,6]{Pedro A.M. Mediano}
\author[1,7,8,9]{Fernando~E.~Rosas}
\author[1,2,14]{Eduardo Viegas}
\author[10,11,12,13]{Pablo A. Marquet}
\author[1,2,14]{Henrik~J.~Jensen}

\affil[1]{Centre for Complexity Science, Imperial College London, London, United Kingdom}
\affil[2]{Department of Mathematics, Imperial College London, London, United Kingdom}
\affil[3]{The Alan Turing Institute, London, United Kingdom}
\affil[4]{Alignment of Complex Systems, Centre for Theoretical Study, Charles University, Prague, Czechia}
\affil[5]{Department of Computing, Imperial College London, London, United Kingdom}
\affil[6]{Department of Psychology, University of Cambridge, Cambridge, United Kingdom}
\affil[7]{Department of Brain Sciences, Imperial College London, London, United Kingdom}
\affil[8]{Department of Informatics, University of Sussex, Brighton, United Kingdom}
\affil[9]{Centre for Eudaimonia and human flourishing, University of Oxford, Oxford, United Kingdom}
\affil[10]{Department of Ecology, Pontificia Universidad Católica de Chile, Santiago, Chile}
\affil[11]{The Santa Fe Institute, 1399 Hyde Park Rd, Santa Fe, NM 87501, USA}
\affil[12]{Centro de Modelamiento Matemático (CMM), Universidad de Chile, International Research Laboratory, 2807, CNRS, CP 8370456 Santiago, Chile.}
\affil[13]{Instituto de Sistemas Complejos de Valparaíso (ISCV), CP 2340000 Valparaíso, Chile}
\affil[14]{Department of Computer Science, Tokyo Institute of Technology, Yokohama, Japan}

\affil[*]{h.rajpal15@imperial.ac.uk}

\begin{abstract}
At what level does selective pressure effectively act? When considering the reproductive dynamics of interacting and mutating agents, it has long been debated whether selection is better understood by focusing on the individual or if hierarchical selection emerges as a consequence of joint adaptation. Despite longstanding efforts in theoretical ecology there is still no consensus on this fundamental issue, most likely due to the difficulty in obtaining adequate data spanning sufficient number of generations and the lack of adequate tools to quantify the effect of hierarchical selection. 
Here we capitalise on recent advances in information-theoretic data analysis to advance this state of affairs by investigating the emergence of high-order structures --- such as groups of species --- in the collective dynamics of the \textit{Tangled Nature} model of evolutionary ecology. Our results show that evolutionary dynamics can lead to clusters of species that act as a selective group, that acquire information-theoretic agency. 
Overall, our results provide quantitative evidence supporting the relevance of high-order structures in evolutionary ecology.
\end{abstract}
\begin{document}

\flushbottom
\maketitle
%
%
\thispagestyle{empty}

\section{Introduction}
The dynamics of life around us is characterised by a plethora of complex interdependent relationships. These relations span across different scales from cells to organisms, and even the biotic and abiotic environment in which they exist. Adaptation through selection is widely agreed to be the motor behind both macro-evolution, as documented in the fossil record~\cite{Gould2002}, and micro-evolution, as e.g. studied in microbial experiments such as~\cite{lenski2004phenotypic}. However, the collective and mutually interdependent nature of evolutionary dynamics raises a critical question: what is the level at which selection \emph{effectively} operates on a set of entangled co-adapting entities?

Since Darwin, the standard view is to assume individual  --- and later genes --- as the drivers of evolutionary change. An extreme version of this view is to regard individual genes as the selective unit, as is often advocated by the so-called `selfish gene' position~\cite{hurst2001role,werren2011selfish} (for a insightful discussion, see Ref.~\cite{Jablonka2005}). Alternative perspectives, where selection also acts at the level of 
higher-order entities (such as groups of species, ecosystems, or even the whole biosphere) on a hierarchical fashion have been the subject of long debates~\cite{Gould2002,Jablonka2005,okasha2006levels,Walsh2015}. Crucially, the possibility that non-reproducing higher-order systems may be subjected to a different kind of selection (e.g. persistence selection~\cite{ford2014natural,doolittle2017s,doolittle2018processes,lenton2021survival}) amounts to a drastic shift in the way we conceive selection and evolution more broadly. Such views raise important questions regarding how group-level dynamics may make the fate of individuals not only depend on their genetic information, but also on how this is integrated into a larger system of interacting entities. Put simply, these views suggest to shift the focus from the singer (i.e. species) to the song~\cite{doolittle2017s} (i.e. relationships among the species).

A way to advance these questions is to avoid reducing them to a dichotomous choice between selection at the basic level of reproduction versus at some higher collective level, and instead consider that different types of selection pressure may be working in tandem at different levels of ecosystemic organisation. However, in order to pursue such a view, it is crucial to have quantitative tools that are capable of identifying and differentiating degrees of (possibly time- and scale-dependent) cooperative selective structure, which could be used to investigate these ideas on empirical or simulated data.

In this paper we address these questions by developing information-theoretic tools to investigate hierarchical selection in data of co-evolving species. Recent theoretical advancements have introduced promising methods based on information storage and predictive information to quantify individuality~\cite{krakauer2020information}. The basic hypothesis behind these approaches is that if a group of individuals can enhance the prediction of their joint future, they can better adapt and thus survive. Building on these ideas, here we present analyses of simulations of co-evolution dynamics based on the well-studied Tangled Nature model \cite{jensen_2018}, and investigate if there are conditions under which selection effectively acts at the level of groups of species instead of single species. 
We explore this question from various complementary angles, including analyses of information-theoretic `individuality'~\cite{krakauer2020information}, integrated information~\cite{balduzzi2008integrated,mediano2022integrated}, and other measures of information dynamics~\cite{lizier2008information,mediano2021towards}. 
Overall, our results provide quantitative evidence suggesting that groups of species act as a unit of selection for biologically plausible mutation rates. Crucially, these higher-order phenomena are observed in the evolutionary dynamics arising from a simple underlying mechanism, which does not includes group level interactions~\cite{gibbs2023can} or interaction delays~\cite{saeedian2022effect}. Thus, these results highlight the spontaneous emergence of groups of cooperating species as a natural consequence of relatively simple processes of adaptation and selection pressure.

\section{Results}\label{sec2}
Our analyses are based on evolutionary trajectories generated by the Tangled Nature (TaNa) model~\cite{jensen_2018}, a well-known computational model that has been extensively used to investigate multiple aspects of evolutionary dynamics, including the observed species abundance curves~\cite{hall2002time}, entropy of species distribution~\cite{roach2017entropy}, hierarchical organization of ecosystems~\cite{becker_sibani_2014}, and the statistics of mass extinctions~\cite{cairoli2014forecasting}. 
The TaNa model establishes the dynamics of the population of multiple species co-evolving over time. In the model, the fitness of each species depends upon the population of species it interacts with as well as the total population of the ecosystem. The results presented in this section are obtained from 10,000 simulations for different values of the mutation rates, and calculating ensemble averages over the results (see \textit{Methods}).

\subsection{Error threshold and population diversity}

As a first step in our analyses, we investigated how the total population and diversity of species is affected by the mutation rate of the evolving agents. 
Special attention is paid to the dynamics observed in the vicinity of the `error threshold,' which is the limit on the mutation rate for species beyond which any biological information needed for continual survival is destroyed in subsequent generation\cite{biebricher2006quasispecies,eigen1988molecular}.

\begin{figure}[h!]
    \centering
    \includegraphics[width=0.7\textwidth]{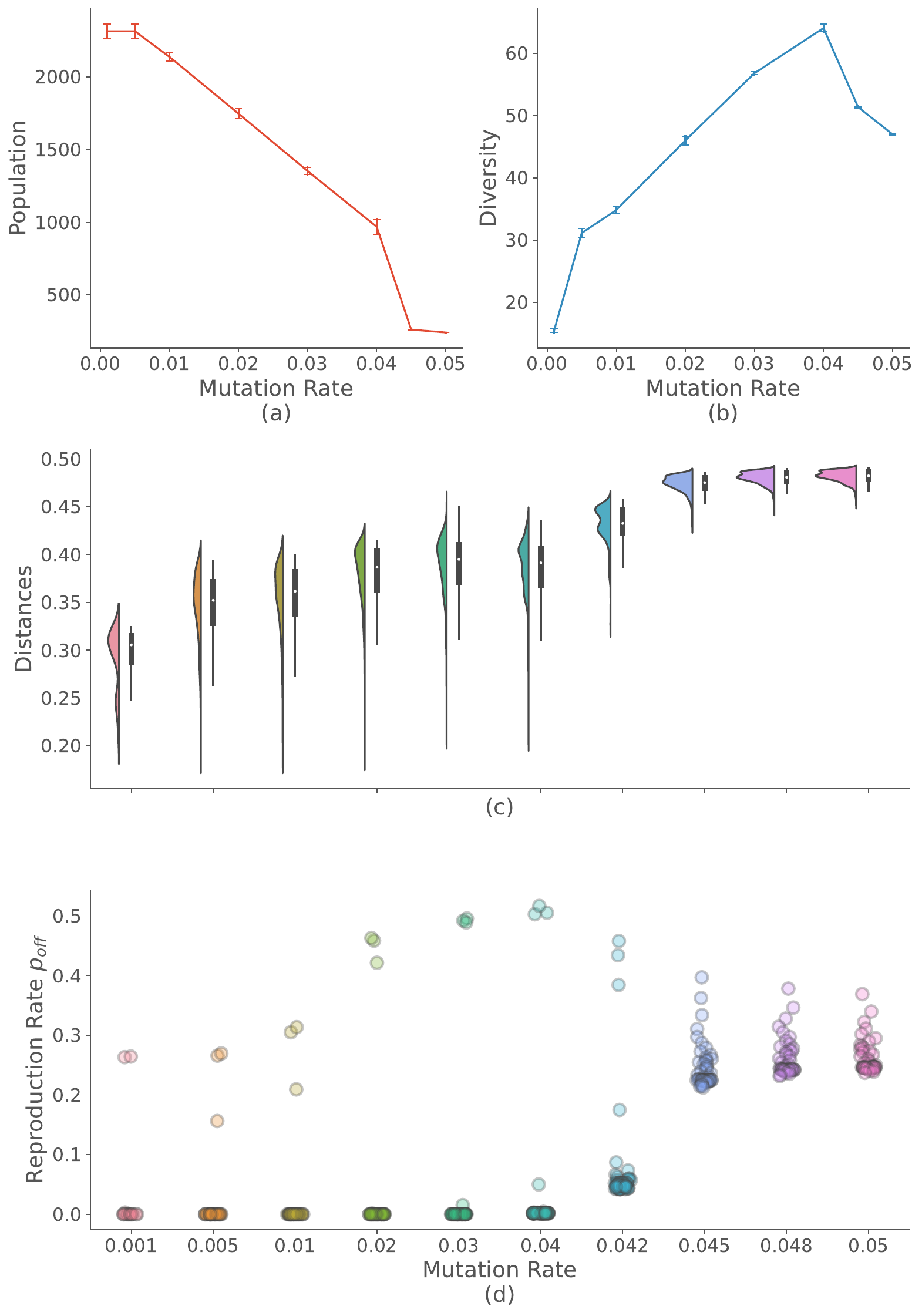}
    \caption{As mutation rate increases, (a) overall population decreases monotonically, while (b) the diversity of species increases until it reaches a peak value for a mutation rate of 0.04. At the same time, with higher mutation rates the existing species of the system (a) become more genetically distant from each other (c), as shown by the distribution of species in the genome space. Figure (d) shows the reproduction probability of each existing species as a scatter plot. It can be seen that for low mutation rates a few core species reproduce at a significantly higher rate than the other existing species.}
    \label{fig:pop_mute}
\end{figure}

As expected, our simulations show that the total population progressively decreases with mutation rate, with a sudden drop after mutation rate 0.04 (see Figure~\ref{fig:pop_mute}). In contrast, the diversity of species increases with mutation rate and peaks at 0.04. This helps us identify the error thresholds associated with the model, beyond which no stable species are observed.

To better understand the effect of the mutation rate, we explored its effect on the average fitness and distance between extant species in the genome space. 
Results show that low mutation rates support the existence of a handful of very fit species with a cloud of mutants around (see Figure~\ref{fig:pop_mute}). In contrast, higher mutation rates allow more non-trivial combinations of species existing near the error-threshold, which is confirmed by the increasing hamming distance among the species and the decreasing gap of fitness between the top few species and the others.

\subsection{Information Individuality}

After identifying the error threshold, we investigated the potential presence of hierarchical selection by estimating the organismal `individuality scores' for different group sizes of species via the framework introduced in Ref.~\cite{krakauer2020information}. 
Briefly, this approach proposes an organismal `individuality score' that represents the degree to which a group behaves as a single entity, in the sense that it is maximally self-predictive (i.e. the group's future evolution is maximally predicted from knowledge of its own past). According to this framework, if a group of species achieves a greater individuality score than a single species, then the group is able to reduce its collective future uncertainty. This enables the group to adapt better and persist for longer as an evolutionary unit, as compared to single species alone.

\begin{figure}[h!]
    \centering
    \includegraphics[width=0.7\textwidth]{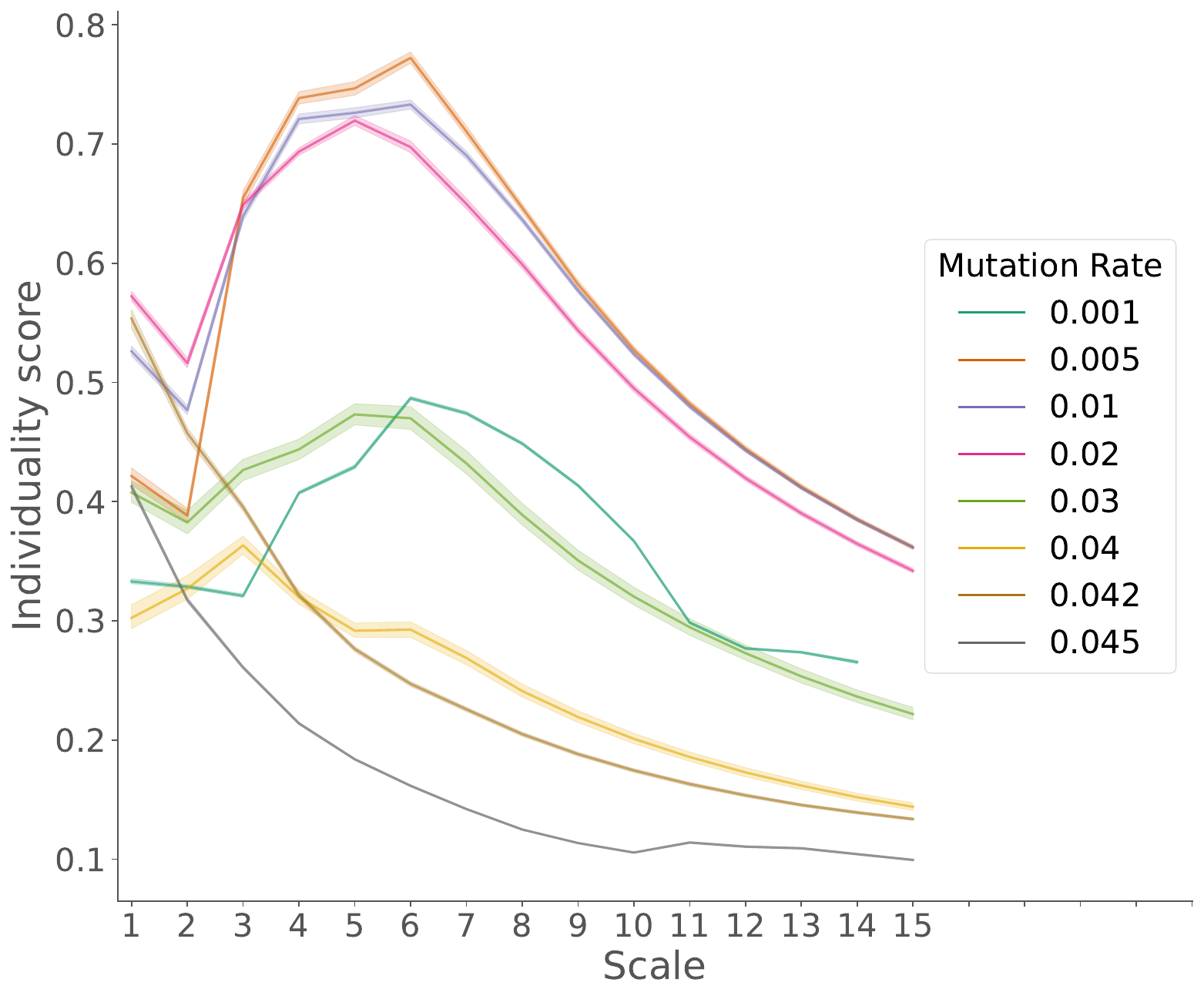}
    \caption{Average organismal individuality scores observed at different scales and mutation rates. A peak of information individuality is observed for intermediate mutation range for higher scales (between 5 and 9). Conversely, in the transition region (mutation rates of 0.042 and 0.045) single species (scale = 1) have the highest individuality scores.}
    \label{fig:individuality}
\end{figure}

Organismal individuality scores were calculated for over 10,000 different combinations for each group size (which we refer to as \textit{scale}) --- singlets (scale = 1), dyads (scale = 2), triplets (scale = 3), and so on. These scores were then normalised based on the size of the group.
Results reveal distinct behaviour as the range of mutation rates changes from below the error transition region to the transition region itself (see Figure~\ref{fig:individuality}). At low mutation rates, organisation into cliques of higher scales becomes apparent and individuality scores peak for scales between 5 and 9. Though the peak starts to flatten out for mutation rates $0.03$ and $0.04$, higher-order organisation still persists. For mutation rates in the transition range, the higher order organisation is lost and single species level becomes the most optimally self-predicting scale.

Formal definitions of the individuality score presented here can be found in section~\ref{sec:info_ind}. These results are also replicated on other proposed measures of individuality, which confirms the presence of higher-order organisation (see Appendix~\ref{appx:col_ind}), irrespective of the measure used.

\begin{figure}[h!]
    \centering
    \includegraphics[width=0.8\textwidth]{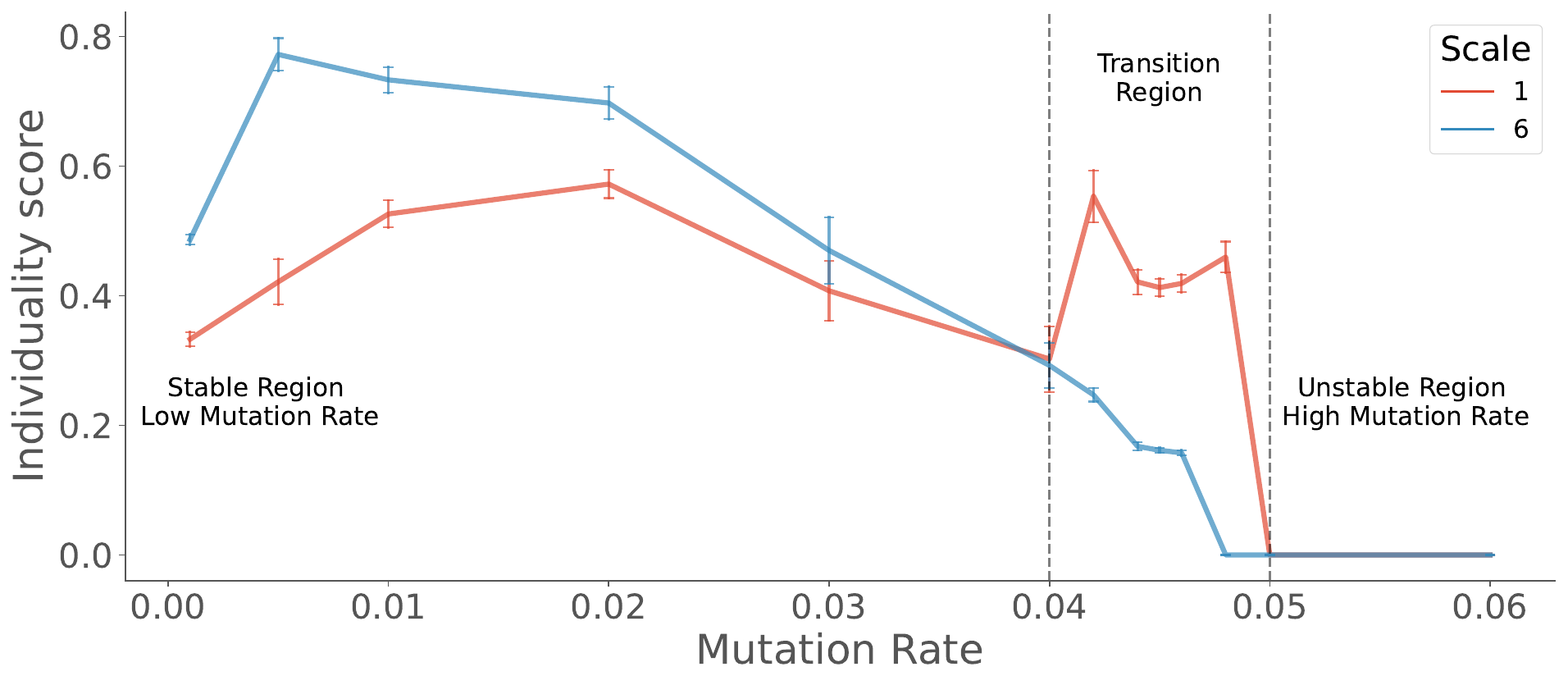}
    \caption{Mean AIS for the two scales of organization -- Individual (Scale 1) and Higher Order (Scale 6) -- varying with mutation rates.}
    \label{fig:ais_16}
\end{figure}

The normalised individuality scores presented here can be interpreted as information carrying capacity~\cite{crutchfield2003regularities}, predictive information~\cite{bialek2001complexity} or information storage~\cite{lizier2012local}. All of these definitions refer to the amount of information in the past of the system that can be used to predict its future. Having identified scale 6 as the optimal scale of organismal individuality we can compare how its individuality changes with mutation rate as compared to single species, i.e. scale 1 (see Figure~\ref{fig:ais_16}).

It can be seen that individuality score of species at scale 6 peaks for low mutation rates and then decreases monotonically through the transition region before going to zero near the error threshold. However, at the scale of single species peak individuality is observed in the transition region. A clear crossover can then be observed in the transition region where higher order organisation looses individuality while the singular species gain organismal individuality. In the following section we explore the species-environment information transfer and integration, to further highlight the role of the error-transition region.

\subsection{Interaction between species and their environment}\label{sec_interaction}
As a last step of our analysis, we characterise the species-environment interactions using measures of information transfer and integration. As discussed before, we compare these interdependencies at the level of single species (i.e. scale 1) against scale 6, which our previous analysis highlighted as the optimal scale of individuality for most intermediate mutation rates. We estimate these measures for various mutation rates in order to identify how a species (or a group of species) interact with the environment near the error threshold.

For the purposes of this analysis, we use the population of a single species (for scale 1), a vector of population of species (for scale 6) or the total population of the environment (all remaining species) as random variables to estimate the information measures discussed below. The population of the environment is calculated as the difference between the total population of the ecosystem and the total population of the species (or the group of species). Further details about these estimations are provided in Section~\ref{sec:info_methods}.

First, we focus on the information transfer as measured by Transfer Entropy (TE)~\cite{schreiber2000measuring} between species and environment. This directed measure estimates the information that a \textit{source} variable about the future state of a \textit{target} variable, over and above the information contained in the past state of the target itself.

\begin{figure}
    \centering
    \includegraphics[width=0.8\textwidth]{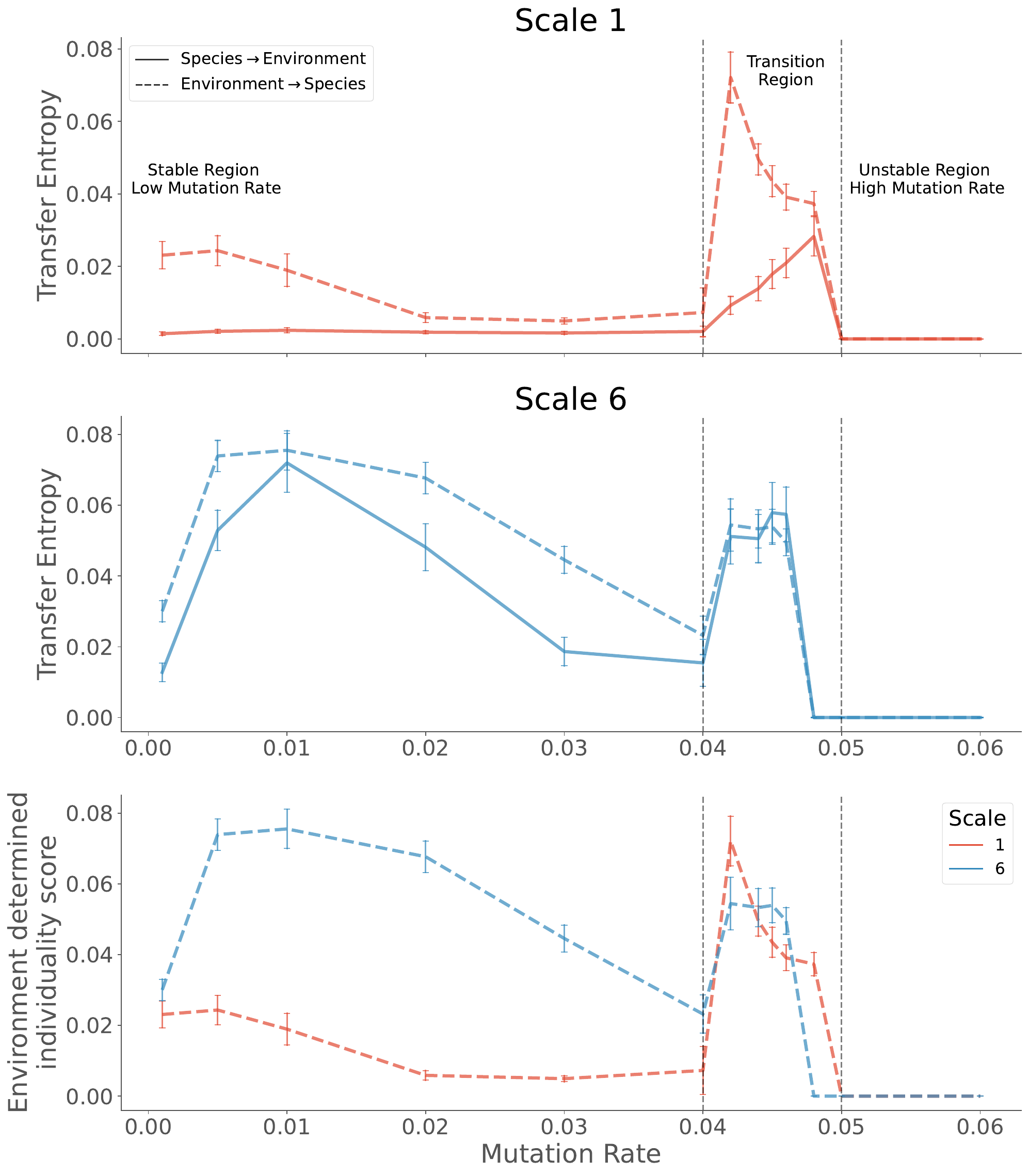}
    \caption{Information flow between species and environment as measured using TE for various mutation rates. The top panel shows the variation at the scale of individual species (scale 1) and the bottom panel for the optimal higher order grouping (scale 6).}
    \label{fig:TE_16}
\end{figure}

The information transfer to and from the environment also varies differently at the two levels of organization (see Figure~\ref{fig:TE_16}). For scale 1, information flow is predominantly from environment to the single species but decreases with increasing mutation rate till the transition region. However, in the transition region information flow peaks in both directions but at either ends of the region.

The trends of information-flow look different for the higher order organisation of species (Figure \ref{fig:TE_16}). For scale 6, there exists a significant amount of information flowing in both directions, though environment to species information flow is larger up-till the transition region. In this case, the information flow peaks in both directions together and stays almost equal throughout the transition region. Thus implying a near symmetric information flow during the error transition.

The transfer entropy from the environment to the group of species, in case of the Tangled Nature model is equivalent to the \textit{Environment determined} individuality~\cite{krakauer2020information}. This measure quantifies how much of the persistence of the group is predicted by the environment beyond what the group can predict. In essence information stored in the environment and its interactions with the group are able to predict the future of the group. When we compare this quantity across the two scales, we can see that in contrast to organismal individuality, higher-order organisation still possesses environment determined individuality in the error-transition region. Whereas the single species exclusively peaks in both individuality scores in the transition region.   

Finally we look at the integrated information as measured using \phiR which captures the average overall coherence among the species and the environment at the single species level (scale 1). 

\begin{figure}
    \centering
    \includegraphics[width=0.8\textwidth]{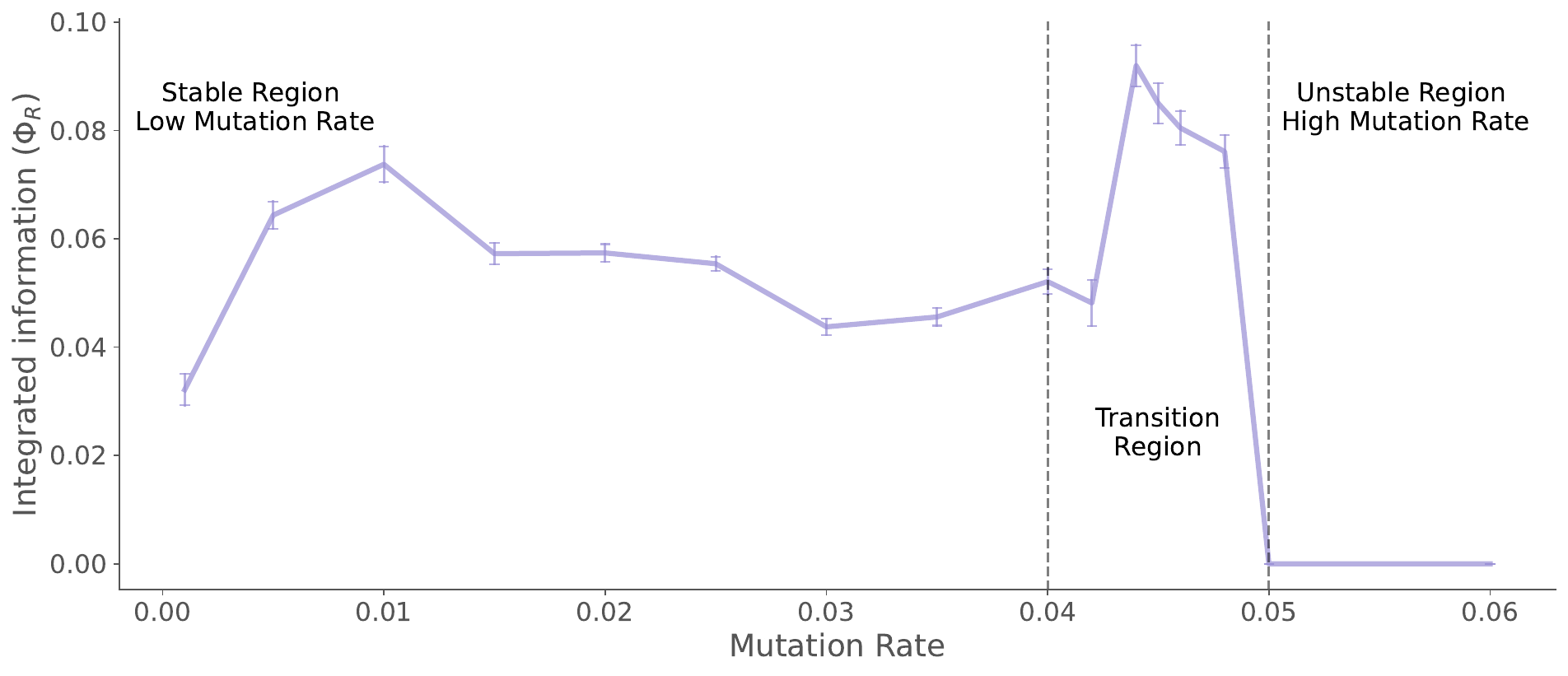}
    \caption{Average Integrated Information between the specie and the environment at the single species level.}
    \label{fig:phi_16}
\end{figure}

Here we also recover a peak of integrated information during the transition region (see Figure~\ref{fig:phi_16}), for scale 1. These peaks in different modes of information processing during an order-disorder transition is consistent with the literature on criticality in complex systems~\cite{lizier2008information,crutchfield1990computation,sole2001information,davis2020phase,mediano2022integrated}.

Overall these results highlight two major findings. First, the species-environment interactions at the level of single species differ from the higher-levels of organisation. Particularly near the error threshold, where the higher-order groups of species loose organismal individuality, while still being environmentally determined. Whereas, at the level of single species, both organismal and environment determined individuality peak during the error-transition region. Which highlights the second finding of this analysis. Peaking of information measures during the error-transitions are in line with the literature on criticality in complex systems~\cite{lizier2008information,crutchfield1990computation}. It suggests that the systems operating close to an order-disorder transition can access many different modes of information processing and thus afford more adaptability and dynamic range. We find that information storage, transfer and integration peaks during this transition at the level of single species, while the higher-order organisation dissolves. In summary, as fluctuations increase, the robustness and resource sharing afforded by the higher-order organisation declines. Therefore, in order to maintain persistence more information is processed at the level of individual species. For instance, RNA viruses are known to operate near the error transition~\cite{domingo2000viruses} for optimal survival.

\section{Discussion}

This paper investigated hierarchical selection in the Tangled Nature (TaNa) model of ecological evolution~\cite{jensen_2018}. 
In contrast to prior work, our work leverages recent advances in information theory to provide analyses that are quantitative and mathematically rigorous, which allows us to objectively estimate the degree of high-order individuality in this self-organising evolutionary system~\cite{krakauer2020information}. 
Crucially, these tools provided evidence of how relatively simple processes of adaptation and selection pressure can result on the emergence of groups of cooperating species that acts as effective units within the evolutionary process. 
Specifically, our results identified signatures of hierarchical selection in the simulated ecosystems, with groups of 5 to 9 species acting as individual evolutionary units. 
Interestingly, the dominance of multi-species evolutionary units breaks down when mutation rates are close to the error threshold, suggesting evolutionarily relevant interactions between hierarchical selection and mutation rates.

The peak in each of the information theoretic measures for groups of 5-8 species at intermediate mutation rates is highly suggestive of hierarchically organised units of selection. In different biological contexts, this could be interpreted as selection of ecological communities \cite{weber2017communityevo}, holobionts \cite{roughgarden2017holobionts}, or viral quasispecies \cite{biebricher2006quasispecies}. The results in this paper suggest that it is highly plausible that emergent units of selection such as these could arise, and the information theoretic measures developed provide a quantitative toolkit for demonstrating this. That said, more work is needed to translate these methods to real-world datasets. 

Clarifying how the proposed measures of hierarchical selection are affected by mutation rates is useful to deepen our understanding of the mechanisms driving ecosystems from an ecological point of view. Increasing mutation rates are generally related, among other things, to harsher environmental conditions or adversity~\cite{wei2022rapid}. Mutations during times of adversity is a strategy to adapt and survive in a changing environment. Therefore, we can understand the mutation rates in the TaNa model as a proxy to varying environmental selection pressure. Error transition -- that marks the limit of mutation rate for existence of stable species -- shows some interesting information processing properties. Firstly, as discussed above the emergent macroscopic organisation breaks down during this region. Simultaneously, individuality scores as well as other information processing measures peak for individual species. Suggesting a trade-off between collective resource sharing (at the macro level) for increased information processing abilities (at the micro level) for continued survival.  

Overall, this work provides an important first step towards enabling quantitative investigations about hierarchical selection, establishing formal methods of analysis that can be readily applied to real data or other models in the future. It is worth emphasising that any ecosystem has a a variety of different species interacting with each other and the environment in unique ways. While the present study focused on average species-environment properties, future investigations could consider more dedicated species-level analyses. 
Finally, another interesting extension of this work could be to apply similar methods to applications of the TaNa model on social scenarios, to investigate if high-order phenomena also take a central role within the dynamics of cultural~\cite{nicholson2016cultural}, organisational~\cite{arthur2017tangled}, and opinion~\cite{rajpal2019tangled} changes.

\section{Methods}\label{sec11}
Here we discuss the details of the model and the information-theoretic measures used for the simulation and subsequent analysis of the model. We provide a brief overview of the Tangled Nature model~\cite{jensen_2018}, followed by the information individuality framework~\cite{krakauer2020information} and other information dynamics measures presented above.

\subsection{The model}

As an agent-based model, species -- represented by a binary genome -- form the dynamical units of Tangled Nature model. These individual species are subject to three stochastic processes: replication, mutations, and annihilation. No groups or hierarchies are defined \textit{a priori}. The evolutionary dynamics takes place in a space of genomes in which random interactions connect different species. The probability that an agent reproduces is determined by a sum over influences from other co-existing species the agent is interacting with. Despite the model being extremely simple, it captures a very broad range of evolutionary phenomena. Starting with a few existing species, the model dynamics evolves to a quasi stable configuration (also known as \textit{evolutionary stable strategies}) where only a select group of species exist. These stable species are disrupted by mutations in the system that drive them to extinction and a new group of species emerges as a result of the reorganization. Over time, the system evolves to more and more stable configurations, thus avoiding these mass extinction events and developing resilience.

Although the model has evolved into different variations over the years, for the purposes of this study we consider the model as defined in the original paper~\cite{christensen2002tangled}. Each species is defined using a unique binary genome of length $L$, comprising an ecosystem of a total of $M = 2^L$ possible species. Interactions between species are encoded in an interaction matrix $J_{M \times M}$. All entries in the interaction matrix are sampled at random from a uniform distribution, $J_{i,j} \sim \mathcal{U}(-1,1)$, thus allowing potentially symbiotic, competitive or predator-prey relationships between any pair of species. However, of all possible interactions, the number of permitted interactions is controlled by coupling probability $\Theta$. The population of a given species at a given time is represented as $n_i(t)$ and the total population of the ecosystem is represented as $N(t)$. 

Starting from a random set of populations of existent species, each timestep starts with an annihilation step where a member of a species, selected uniformly at random, is killed with a probability $p_{\text{kill}}$. This is followed by asexual reproduction, where a member of a species, selected uniformly at random, creates an offspring with probability $p_{\text{off}}$. Each gene of the created offspring's genome undergoes a mutation with a probability $p_{\text{mut}}$. These mutations introduce new species in the ecosystem which then compete with the existing species. The state of the system is recorded after each generation, $N(t)/p_\text{kill}$ timesteps, which is the average number of timesteps required to kill all currently existing species. The reproduction probability $p_\text{off}$ depends upon the fitness of the species at the given timestep. The fitness function, which is a weighted sum of interactions with all other species, is defined as
\begin{equation}
	\centering
	\mathcal{H}(n_i,t) = \frac{k}{N(t)}\sum_{j=1}^M J_{i,j} n_j(t) - \mu N(t) ~ ,
	\label{fitness}
\end{equation}
\noindent where $\mu$ represents the carrier capacity or the resource constraints driven by increasing population, and which has a negative contribution to the fitness. Whereas, $k$ is a scaling parameter for the strength of the couplings. Thus, the fitness of a given species depends not only on how it interacts with other neighbouring species but with the rest of the environment as well. The fitness function is related to the reproduction probability $p_\text{off}$ for a given species $i$ at timestep $t$ as
\begin{equation}
	\centering
	p_{\text{off}}(n_i,t) = \frac{1}{1 + \exp^{-\mathcal{H}(n_i,t)}} ~ .
	\label{prob_offspring}
\end{equation}

Note that the probability of reproduction is non-linearly related to the fitness function. Although the probability of reprodution is higher for species with positive fitness, some non-zero probability of reproduction exists for negative fitness values, which enables non-performing species to reproduce and mutate towards fitter species.

\begin{figure}[h]
	\centering
	\includegraphics[width=\textwidth]{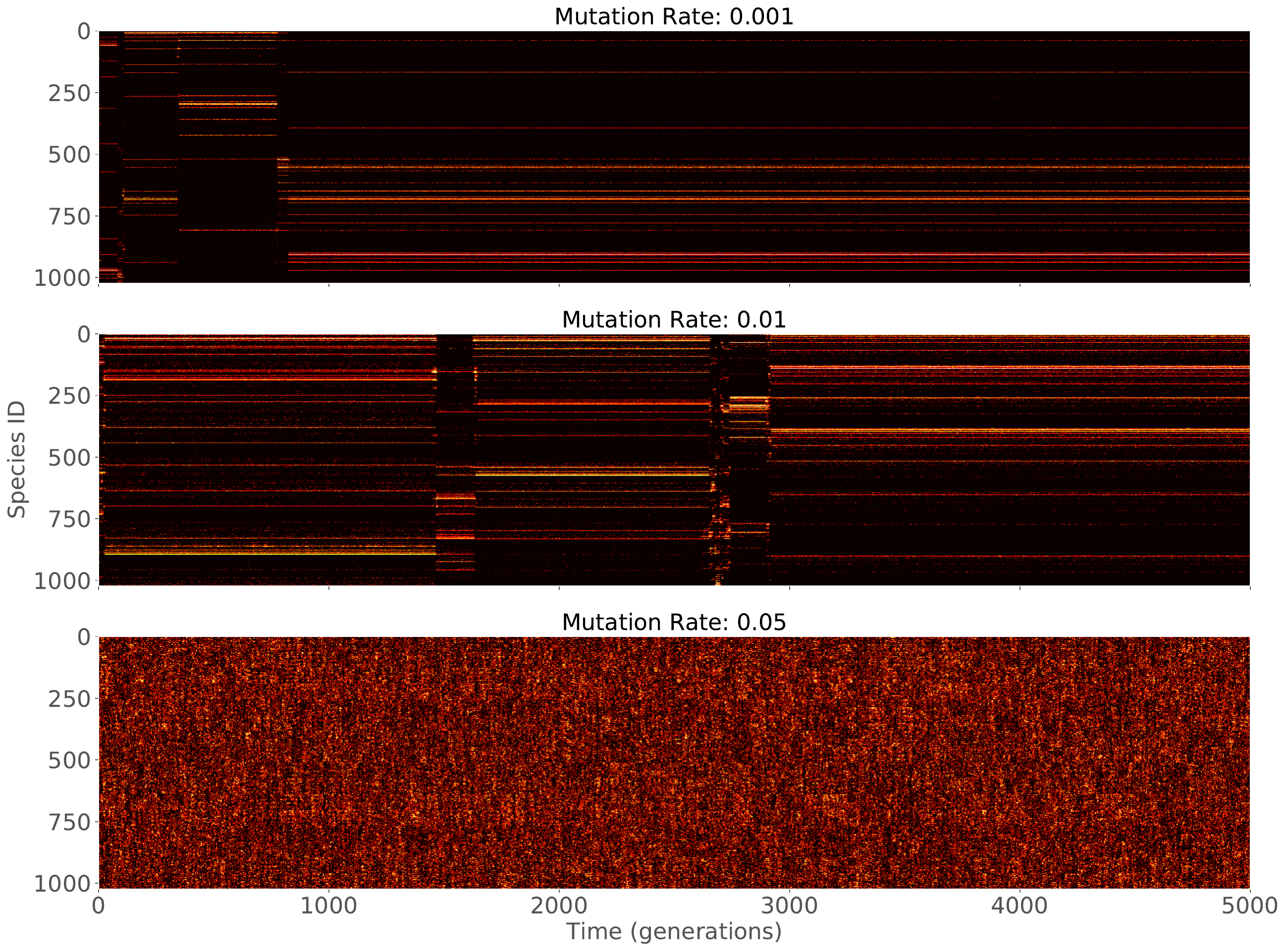}
	\caption[Existing species for different mutation rates]{The figure shows TaNa evolution recorded for 5000 generations, for standard parameters $L = 10$, $\Theta = 0.25$, $p_{\text{kill}} = 0.2$, $w = 33$ and $R = 143$, for three different mutation rates. The bright dots indicate the existing species out of the $2^{10}$ available species. Stable q-ESS are present for $p_{\text{mut}} < 0.05$. }
\label{mutation_tana}
\end{figure}

For the purposes of our study, the fixed parameters used for the model are, $L = 10$, $\Theta = 0.25$, $p_{\text{kill}} = 0.2$, $w = 33$ and $\mu = 1/143$. These parameters are chosen based on the standard parameter ranges used in previous studies~\cite{christensen2002tangled}. We study the changes observed in the dynamics of the model when a key parameter $p_{\text{mut}}$ is varied. This parameter represents the selection pressure introduced by the ecosystem. Since changing environmental conditions lead to more mutations, $p_{\text{mut}}$ can be considered as a proxy for controlling environmental selection pressures~\cite{wei2022rapid}. This parameter has significant impact on the dynamics of the system: visually, it can be observed (see Figure~\ref{mutation_tana}) that the dynamics is more selective and stable with fewer transitions at very low mutation rates ($p_{\text{mut}} = 0.001$). In an intermediate range ($p_{\text{mut}} = 0.01$), more species are observed during the intermittent stable states or the q-ESSs (quasi evolutionary stable states), along with more transitions. Finally, for very high mutation rates ($p_{\text{mut}} \geq 0.05$), new species emerge and old species die every generation and q-ESS are non-existent (i.e. no stable species emerge). Thus, varying the mutation rate provides two interesting transition points: a first one where more interesting combinations of species start to emerge as we move from very low to intermediate range; and a second transition where the system moves from order to disorder between the range of $p_{\text{mut}} \in (0.4,0.5)$. 

Computationally efficient Rust code was used to simulate the Tangled Nature Model. The code is available with documentation on \href{https://github.com/clem-acs/tangled-nature}{GitHub}.

\subsection{Information-theoretic measures}\label{sec:info_methods}

In this paper we use tools from information-theory to estimate the various measures presented in the Results. Primarily, we use multivariate mutual information (MI) to quantify interdependencies between time series of species populations obtained from the simulations. Typically, numerical estimation of MI requires stationary probability distributions of the random variables. Since the tangled nature model exhibits non-stationary evolution, we use ensembles of simulations to estimate mutual information. Details of this ensemble method of estimating mutual information is provided in Appendix~\ref{appx:ensemble}. Below, we briefly discuss the measure of information individuality, as well as other measures used to quantify species-environment interactions. 

\subsubsection{Information individuality}\label{sec:info_ind}
 
In recent work, Krakauer and others\cite{krakauer2020information} put forward an information-theoretic solution to identifying the boundary between an individual and its environment. This definition is based in principles of optimal self-prediction --- i.e. if a subsystem can predict its future better than any of its parts, and any addition to the subsystem hinders its predictability, then that subsystem is deemed an information individual. This optimal self-predictability enables biological entities to control and navigate their environment~\cite{hoel2020emergence}, implying that selection for persistence~\cite{lenton2021survival,doolittle2018processes} could be a putative explanation for their emergence. 

For the analyses above, let $\mathbf{S}(t)$ be a joint vector representing the population of a subset of $K$ species, $\mathbf{S}(t) = (n_1(t), n_2(t),\dots,n_K(t))$, at a given time $t$. If $N(t)$ is the total population of the ecosystem, the corresponding environment $E(t)$ can be then written as 
\begin{equation}
    E(t) = N(t) - \sum_{i=1}^K n_i(t) ~ .
\end{equation}

Then, based on the properties laid out in the original paper~\cite{krakauer2020information}, Krakauer \textit{et al.} write three different individuality measures as follows:

\begin{equation*}\label{eq:ind_def}
\begin{split}
	\text{Organismal Individuality}\;		A^*  &=  I(\mathbf{S}(t);\mathbf{S}(t+1)) ~ ,\\
	\text{Colonial Individuality}\;		A\hphantom{^*} &= I(\mathbf{S}(t);\mathbf{S}(t+1) \mid E(t)) ~ ,\\
	\text{Environmental determined Individuality}\;	nC &= I(E(t);\mathbf{S}(t+1) \mid \mathbf{S}(t)) ~ .
\end{split}
\end{equation*}

Here we focus on the organismal individuality which links closely in definition with persistence selection~\cite{lenton2021survival,doolittle2018processes}. This individuality measure includes both the collective predictive information of the group of species, as well as the redundancy they share with the environment (see~\cite{krakauer2020information} for more details). Such information is useful for the species to share resources as well as respond to the environment they live in. 

For our analyses, we first compute an ensemble of Tangled Nature simulations as described in Appendix~\ref{appx:ensemble}. We then randomly sample many different subsets of $K$ species, and take the average organismal individuality of each subset by estimating the mutual information between the current and future populations of the species. This process is repeated for a $K$ values ranging from $K=1$ to 15. To account for the bias that is introduced by increasing dimensions as we calculate multivariate mutual information, we normalize using the size of the group. Thus, the normalized individuality score as shown in Figure~\ref{fig:individuality} can be written as
\begin{equation}
\text{Individuality score} = \frac{I(\mathbf{S}(t);\mathbf{S}(t+1))}{K} ~ .
\end{equation}

\subsubsection{Information transfer and integration}

Finally, we briefly describe the measures shown in Section~\ref{sec_interaction} to quantify species-environment interaction. We keep the same notation as above, using $\mathbf{S}(t)$ to denote the vector representing the population of $K$ species at time $t$, and $E(t)$ to denote the state of ther environment at time $t$.




Transfer Entropy (TE) is a conditional mutual information (CMI) based measure of Granger causality. TE quantifies information transfer from a source variable to the target as CMI between the past of the source and the future of the target conditioned on the past of the target. For instance, TE from the a group of $K$ species $\mathbf{S}$ to their environment $E$ can be written under the Markov condition as,
\begin{equation}
    \text{TE}(\mathbf{S} \rightarrow E) = I(\mathbf{S}(t);E(t+1) \mid E(t))
\end{equation}

Measures of integrated information (generally denoted by $\Phi$), were first introduced by Tononi \textit{et al.}~\cite{tononi1994measure} to measure integration among different regions in the brain. Since then, multiple related measures have been proposed, with some adapted to more practical scenarios~\cite{balduzzi2008integrated,barrett2011practical} and applied to quantify interactions across a broad range of complex systems~\cite{mediano2022integrated}. In essence, these measures quantify the extent to which the interactions between parts of a system drive the joint temporal evolution of the system as a whole --- a system has high integrated information if its dynamics strongly depend on the interactions between its parts.

Here, we estimate two measures of integrated information (whole-minus-sum integrated information, \wmsphi~\cite{balduzzi2008integrated}, and its revised version, \phiR~\cite{mediano2019beyond}) between a single species and its environment jointly evolving over time. Denoting the population of species $i$ and its environment $E$ by the joint random variable $\mathbf{X} = (n_i, E)$, \wmsphi is given by
\begin{equation}
    \centering
    \wmsphi = I(\mathbf{X}(t); \mathbf{X}(t+1)) - \sum_{i=1}^2 I(X_i(t); X_i(t+1)) ~ ,
    \label{eq:phi_tononi}
\end{equation}
\noindent where $X_i$ denotes the $i$\textsuperscript{th} element of $\mathbf{X}$.



Despite its intuitive formulation, \wmsphi as defined above has one important disadvantage: it can become negative in systems where the parts are highly correlated~\cite{mediano2018measuring,mediano2019beyond}. To address this problem, Mediano \textit{et al.} proposed a revised measure of integrated information, \phiR, based on the mathematical framework of integrated information decomposition ($\Phi$ID)~\cite{mediano2019beyond}. This revised measure simply adds a new term to \wmsphi correcting for the correlation, or redundancy~\cite{williams2010nonnegative}, between the parts of the system:
\begin{equation}
    \centering
    \phiR = \wmsphi + \min_{i,j} I(X_i(t); X_j(t+1)) ~ .
    \label{eq:phi_R}
\end{equation}


In the main text we report results using \phiR, due to its better interpretability. For completeness, we provide a comparison between the two measures of integrated information in Appendix~\ref{fig:phi_comp}.

\section*{Acknowledgments}
H.R. and H.J.J are supported by the Statistical Physics of Cognition project funded by the EPSRC (Grant No. EP/W024020/1). H.R. is also supported by the Ecosystem Leadership Award under the EPSRC Grant EP/X03870X/1; the Economic and Social Research Council under Grant ES/T005319/2;  \& The Alan Turing Institute. F.R. was supported by the Fellowship Programme of the Institute of Cultural and Creative Industries of the University of Kent, and the DIEP visitors programme at the University of Amsterdam. P.A.M. acknowledges support from grants Fondecyt 1200925, Proyecto Exploración 13220184 and through Centro de Modelamiento Matemático (CMM), Grant FB210005, BASAL funds for Centers of Excellence from ANID-Chile. HJJ also thanks EPSRC for supporting this work as part of the Quantifying Agency in time evolving complex systems project (Grant no. EP/W007142/1).

\bibliography{references}
\appendix
\renewcommand\thefigure{\thesection.\arabic{figure}}
\newpage
\section*{Supporting Information}

\section{Alternative individuality scores}\label{appx:col_ind}
\setcounter{figure}{0} 
As mentioned in the paper, we focus on on the organismal individuality as defined in~\cite{krakauer2020information}. However, we replicated the analysis for the colonial individuality scores and found very similar hierarchical organization as reported in the results above.

\begin{figure}[h!]
    \centering
    \includegraphics[width=0.7\textwidth]{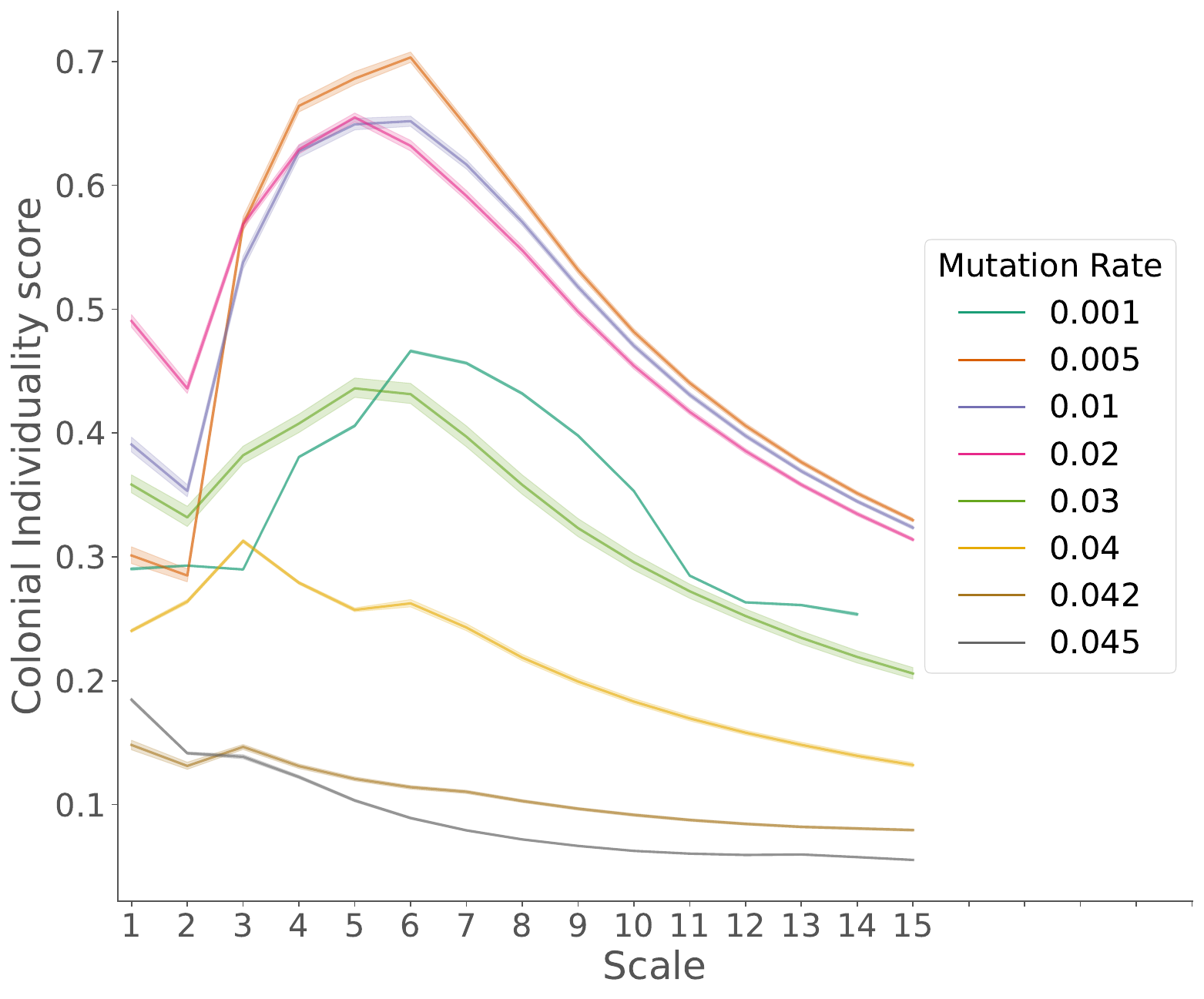}
    \caption{Colonial individuality scores for different scales of organization and mutation rates.}
    \label{fig:col_ind}
\end{figure}

\begin{figure}[h!]
    \centering
    \includegraphics[width=0.7\textwidth]{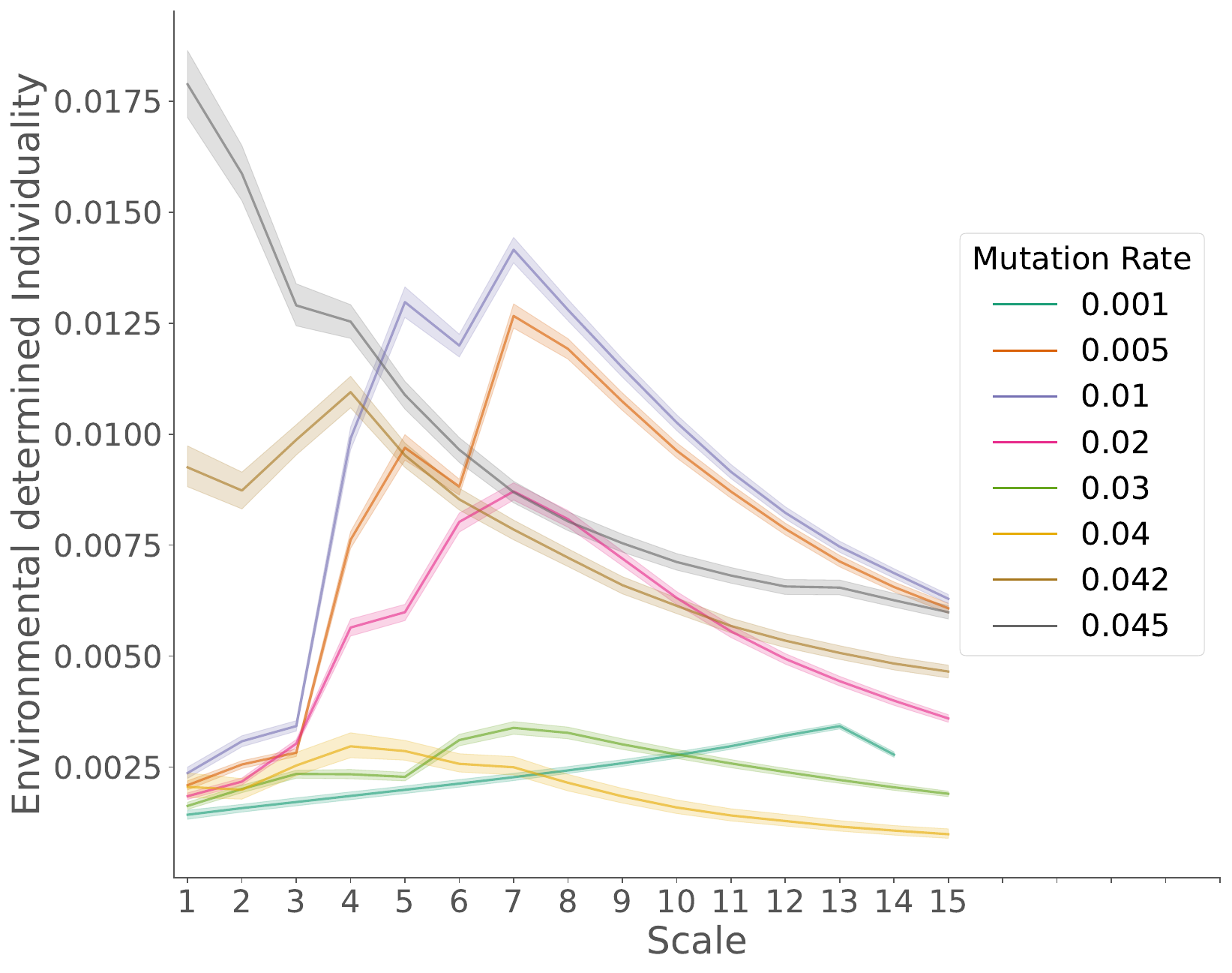}
    \caption{Environment dependent individuality scores for different scales of organization and mutation rates.}
    \label{fig:env_ind}
\end{figure}

As seen in figure~\ref{fig:col_ind} and ~\ref{fig:env_ind}, both colonial and environment determined individuality scores exhibits higher individuality scores for scales 5-9 and intermediate mutation rates. This is in line with the results presented for the organismal individuality in the main results~\ref{fig:individuality}. However this peak flattens as we approach error-transition. Near this threshold, individual species have the highest colonial and environment dependent individuality scores.

\section{Non-Stationary dynamics}\label{appx:ensemble}
The tangled nature model exhibits a time dependent non-stationary dynamics. Such that the inter-transition intervals decay slowly in time. As discussed before the system evolves into more and more stable configurations over time so the number of mass-extinction events goes down in time. The population and diversity also varies logarithmically in generational time. Therefore, the probability distribution of any given species or the whole population changes with time. Therefore, to estimate information-theoretic measures, we estimate probability distributions for each timestep across an ensemble of simulations. For the current analysis we use over 10,000 simulations for every mutation rate to estimate information-theoretic measures. In order to study long term behaviour we sub-sampled the data and calculated the measures for every 1000th and 1001th generations. This enables us to calculate information measures like Transfer entropy and Information storage with time delay 1 for every 1000th generation. Having such an ensemble of time-series enables us to estimate information measures for non-stationary time-series.

\section{Integrated information decomposition}\label{appx:integrated}
\setcounter{figure}{0} 
Recently, the PID formulation for single target (see \cite{williams2010nonnegative}) was extended to multiple targets under the Integrated Information Decomposition ($\Phi$ID)~\cite{mediano2019beyond}. Under this extended framework, redundant, unique and synergistic information shared between multiple sources and targets can be separately defined. Thus for a system with two parts, 16 information atoms are now defined for respective source and target. For example, $\text{Unq}^1 \rightarrow \text{Unq}^2$ shows unique information transferred from variable 1 in the source to variable 2 in the target. Similarly $\text{Unq}^2 \rightarrow \text{Syn}$ shows the information that was unique to variable 2 in the source and is now synergistically available in the target. Exact mathematical definitions of these atoms can be found in the $\Phi$ID paper~\cite{mediano2019beyond}.

This new framework of quantifying temporal information relationships in a system also helps in addressing some of the issues with the $\Phi$ measure of functional segregation and integration discussed above(see equation\ref{eq:phi_tononi}). Firstly, the original $\Phi$ measure provided the same numerical value for fundamentally different information processes. By decomposing $\Phi$ into information atoms, these processes can now be differentiated. This decomposition can be written as follows,

\begin{equation}
\begin{split}
	\Phi &= \text{Syn}\rightarrow\text{Syn}\: +\: \text{Syn}\rightarrow\text{Unq}^i\: +\: \text{Syn}\rightarrow\text{Red} \\
 	   	 &+\: \text{Unq}^i\rightarrow\text{Syn}\: +\: \text{Red}\rightarrow\text{Syn}\: +\: \text{Unq}^i\rightarrow\text{Unq}^j \\ 
 	   	 &-\: \text{Red}\rightarrow\text{Red}
\end{split}
\end{equation}

Secondly, the negative double redundancy term ($\text{Red}\rightarrow\text{Red}$), yields $\Phi$ negative for systems with a lot of redundant information flow. So by adding back this term and keeping the synergistic and transfer terms intact, a non-negative measure of information integration $\Phi_R$ can be defined\cite{luppi2021like},

\begin{equation}
\begin{split}
	\Phi &= \text{Synergy} + \text{Transfer} - \text{Redundancy} \\
	\Phi_R &= \Phi + \text{Redundancy} \\
	\Phi_R &= \Phi + \text{Red}\rightarrow\text{Red}
\end{split}
\end{equation}

\begin{figure}
    \centering
    \includegraphics[width=0.65\textwidth]{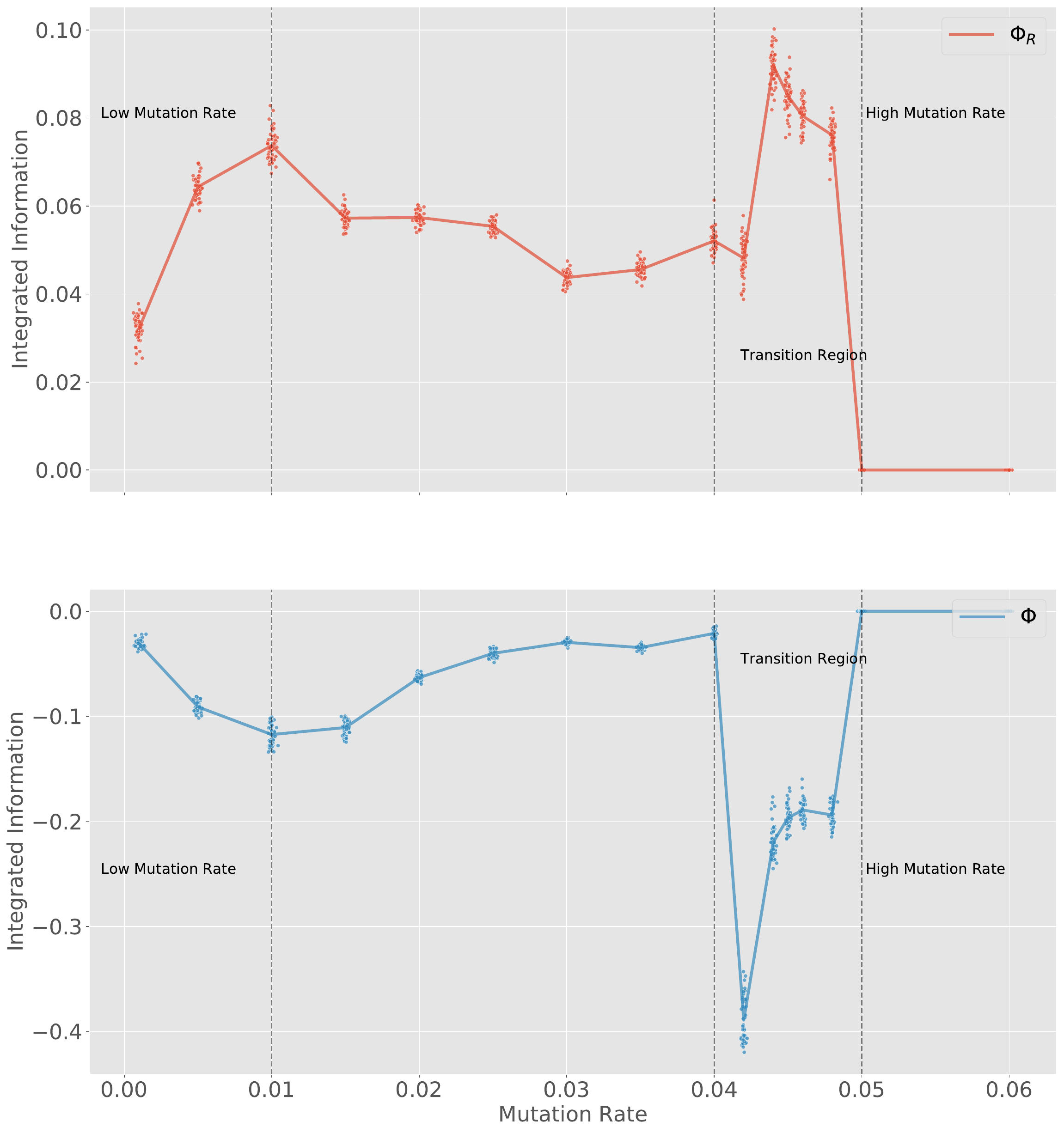}
    \caption{Comparison of $Phi_R$ and $Phi$ measures of integrated information for varying mutation rates.}
    \label{fig:phi_comp}
\end{figure}

Systems with high redundancy among the parts often show negative values for $\Phi$. This is relevant for the Tangled Nature model as evolutionary models often have a high degree of redundancy. We found this to be true in the case of TaNa model, as seen from the figure below~\ref{fig:phi_comp}. 

It can be seen in the figure~\ref{fig:phi_comp} that where a positive peak is observed for $\Phi_R$ during the error-transition region, $Phi$ shows a negative peak. Such a negative peak would imply lack of integration between the species and the environment. However, $\Phi_R$ confirms the contrary.

\end{document}